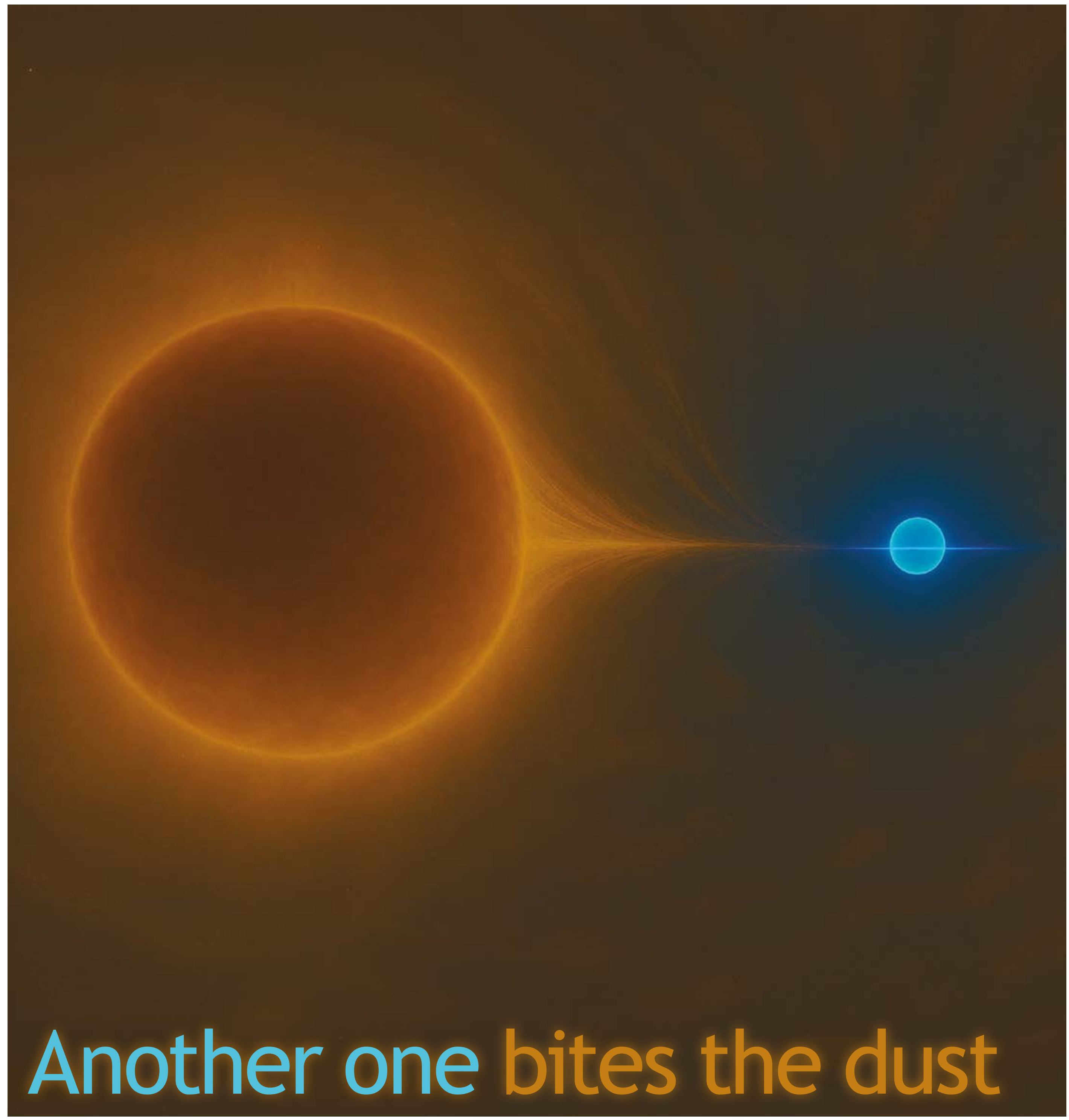

# Another one bites the dust

Extreme red supergiant WOH G64 has led researchers a merry dance these last few years, but what does it reveal about the secret lives of these beasts? Jacco van Loon explains

1 AI impression of the WOH G64 system, composed of a red supergiant and a smaller hot star. New observations with SALT suggest that the hot star may be causing the atmosphere of the red supergiant to spread out. Some of it may fuel a disc of gas around the hot star, giving rise to emission lines in the SALT spectrum (not to scale). (Jacco van Loon/Copilot)

Astronomers are currently faced with a bonanza of red supergiant revelations, from their links to supernovae to their weird and wondrous quirks that catch us out but open avenues to deep insight. One such case has been part of my entire professional astronomical career: the extreme red supergiant WOH G64. It's taken us by surprise in recent years and I can't wait to find out what it's got up its sleeve. Here's why I am so excited.

## Setting the stage

I always tell people that stars are like us: they are born, they live, and then they die. Both stars and humans are transient, unsustainable, irreversible, out-of-equilibrium systems that serve for the universe to move energy through with the overarching goal of increasing entropy. And the death of a star – like human death – is not just catastrophic for the star, but also has a tremendous impact on its surroundings, too. While we witness it happening all too often, it remains a dramatic and mysterious event.

Most massive stars, in the approximate birth-mass range 8–30 solar masses, swell up once helium burning in the core commences. Their luminosity changes relatively little, their surface temperature drops from 20000 or 30000K to approximately 4000K or cooler still. This makes them swell from diameters maybe ten times that of the Sun, to sizes matching the orbits of solar system planets.

They become red supergiants and may easily live on for several hundred thousand years, before the core collapses and – in many cases at least – a supernova ensues, launching the stellar mantle into space and processing it with a neutron-laden blast. This has been the standard picture for many decades, and we wouldn't expect to get any advanced warning of such a supernova: it kept its secrets deep inside. Indeed, the vast majority of supernova progenitors that have been identified are red supergiants, responsible for various types of hydrogen-rich supernovae.

This picture is in the process of being redrawn. The first alteration came in 1987, with the discovery of a supernova in the Large Magellanic Cloud (LMC) galaxy, a 'large dwarf' galaxy currently cruising through the Milky Way halo at a distance of around 160 000 lightyears. The progenitor star was quickly found on historical photographic plates and it had been a blue supergiant. Whatever happened there? The Hubble Space Telescope, once its eyesight had been restored, spotted a ring around the exploded blue star, which further observations from the ground revealed is dusty and was most likely produced when the star was a red supergiant. The star must have transformed itself some 10 000 years before it finally died. But how, and why?

**"The star must have transformed itself some 10 000 years before it finally died. But how, and why?"**

Red supergiants run a race against the clock, blowing a dense wind that gradually strips them of mass at rates that see a Sun's mass being lost in a million years or less – sometimes much less (van Loon 2025). If they lose their mantle before the core collapses, they will no longer be a red supergiant when they explode. The proposal that has gained most traction to explain SN 1987A, however, involves a companion star, with which it merged. The merger or preceding interaction removed much of the red supergiant's envelope, with the binary gravitational configuration breaking the spherical symmetry needed to explain it ending up in a ring. Averaged over time, two stars in orbit around one another define an axis (through the centre-of-mass) and a plane (through the orbits) of symmetry where matter is expected to accumulate. The binary will have existed within the confines of the ring.

Bring on Betelgeuse ($\alpha$ Orionis). Thanks to Arabic scholars we have this red supergiant with the cutest of names, at least for anglophiles: 'Beetlejuice'. More seriously, it is one of the closest red supergiants (if at a safe distance of around 600 light years) and its surface has even been imaged. Unsurprisingly, it is a middle-of-the-road red supergiant, pretty luminous but not terribly cool, of early-M spectral type and thus boasting moderately strong molecular absorption. Consequently, it has not enveloped itself in a thick dust shell as much cooler red supergiants such as VY Canis Majoris and NML Cygni do. It varies in brightness, in a semi-regular way with a quasi-period just over a year (due to large convection-induced surface variations) and another period of about six years. Skilled amateur astronomers can discern these modulations and the American Association of Variable Star Observers database remains the authoritative account of Betelgeuse's behaviour over the past century.

As if heralding imminent doom and disaster on Earth, around the change of year from 2019 to 2020 it had become clear that Betelgeuse had faded more than ever recorded before. The 'Great Dimming' was by some taken as a sign of a lurking core collapse, but we now know that was a storm in a teacup and a much more banal explanation is more likely. One of the convection cells breaking at the surface must have launched a particularly dense parcel of gas, which then traversed above a dark, retreating cell instantiating proficient dust condensation. This dust cloud then transited across the surface facing us (Montargès *et al.* 2021). These dust bunnies had already been imaged around Betelgeuse in infrared light, possibly testifying to similar occurrences in the centuries gone by. This scenario relies on some transverse motion, in line with the remarkably rapid rotation of Betelgeuse. Both the rotation and six-year variation in brightness have been taken as strong evidence for a binary companion having induced these peculiarities. While an earlier sighting met with scepticism, the companion was spotted again in 2025 (Howell 2025).

Interpreting individual events always must be done with caution and within the context of the prevailing general understanding, but error, bias or fashion can still play a part. Over the past two decades it has become appreciated that the majority of massive stars have a massive companion, often close enough to result in interaction either early on or in later life when one of them increases in size. The interaction can lead to a merger, which is the currently favoured explanation for the bizarre eruption of V838 Monocerotis and other red novae. It can also lead to the stripping of the red supergiant envelope, laying bare a compact, hot star that would explode as a hydrogen-deprived supernova. The progenitor of SN 1987A, Betelgeuse, and its rival closest red supergiant Antares ($\alpha$ Scorpii) and other examples such as VV Cephei fit the picture of red supergiant binarity (though in the case of Antares the separation is rather large, at more than 500au). The excitement accompanying Betelgeuse's Great Dimming was in part also incited by the discovery of eruptive behaviour, in a few cases, and much elevated mass loss in a growing number of cases, in the years leading up to a recorded supernova (Elias-Rosa *et al.* 2024). Although the enhanced mass loss could be understood as an extended atmosphere from which only a small fraction is actually lost, and eruptions could be luminous blue variable behaviour.

## Enter the monster

WOH G64 is one of the largest stars known, and a veritable monster. It entered the astronomical world in modest disguise, as an apparently ordinary red giant star in the LMC discovered by Westerlund, Olander and Hedin in the 1960s (Westerlund *et al.* 1981). But once Elias, Frogel and Schwering identified it as the counterpart of the bright mid-infrared source IRAS 04553-6825 it suddenly rose to the status of luminous, dusty red supergiant (Elias *et al.* 1986). That arose plenty of interest and it became the principal target for new observations, such as the discovery of the first extragalactic circumstellar masers, OH, SiO and water (van Loon *et al.* 2001). Monitoring in the optical and infrared exposed a rather regular, large amplitude variation, more like pulsating Mira variables and quite extraordinary for luminous supergiants, that usually display smaller and more irregular variations. The long period of around 900 days was commensurate with its M7 spectral type and high luminosity of about 500 000 solar luminosities. The masers traced an accelerating wind, presumed to be driven by radiation pressure onto dust grains that form at around ten stellar radii distance. It seemed to be a poster child of extreme red supergiant-ness: the biggest star, in the LMC at least, measuring 1500–1700 times the size of the Sun (Levesque *et al.* 2009), encompassing the orbit of Jupiter and then some. And we thought we wouldn't learn much more about it in our lifetime. How wrong we were.

## Changing views

In 2020, Keiichi Ohnaka used the upgraded Very Large Telescope Interferometer at the European Southern Observatory (ESO) in Chile to discover that WOH G64 had blown a dense dust cocoon, probably sometime in the 2010s (Ohnaka *et al.* 2024). Examining other available data in an attempt to make sense of this surprise, he found that the behaviour of WOH G64 had changed dramatically over the course of this millennium. The star had faded and the pulsations had diminished. Around the same time, a Greek-led team announced on ArXiv that the star had changed into a yellow hypergiant, perhaps heralding the loss of much of its hydrogen-rich envelope. They had found an X-Shooter spectrum in the ESO archive, taken by my former PhD student Steven Goldman who had not had time to inspect it given the vagaries of job hunting. It showed a host of emission lines from neutral and singly ionised atoms, betraying the presence of a hot component. Meanwhile, I had been using my access to the Southern African Large Telescope (SALT) to take spectra that I hoped would explain what Keiichi had found. The SALT spectra not only confirm the emission lines, but they also show unmistakable absorption from titanium oxide molecules arising in a cool atmosphere like that of a red supergiant (van Loon & Ohnaka 2026). It looked more like a semi-transparent, maybe patchy, layer overlain on a warmer stellar mantle, but it didn't look quite right.

Now we had known there was something odd with WOH G64 all along. Already in the 1980s the optical spectrum displayed some of the emission lines reported more recently, and they have been seen ever since. WOH G64 has a hot companion star. The spectral energy distribution, first, and maser profiles, next, strongly implicated the companion in shaping the dusty wind such that we have enjoyed a relatively transparent window into the central star despite its thick dust envelope. Could that companion star have anything to do with the mesmerising show WOH G64 put on recently?

When we wrote our letter to *Monthly Notices of the Royal Astronomical Society* in December 2025, presenting our SALT spectra from 2024 and 2025 (van Loon & Ohnaka 2026), we proposed a scenario that might explain what we observed. If the orbit of the hot companion star is substantially elliptical, it may have been the case that it was on the approach around the year 2000. From then on, the pulsation period seems to have shortened slightly and the 'surface' temperature to have increased a little. At the time this was nothing to be too perplexed about, but it may have signalled the beginning of a gravitational interaction between the two components.

The shallower gravitational potential well in the already tenuous atmosphere of WOH G64 A (the red supergiant) resulting from the potential well of the hot star (WOH G64 B) might have allowed the atmosphere to expand even more, becoming partially transparent. This would render a decreased size of the optical continuum source, leading to an earlier spectral type and shorter pulsation period, while still exhibiting some, but shallow, absorption from TiO (and vanadium oxide). The brightness

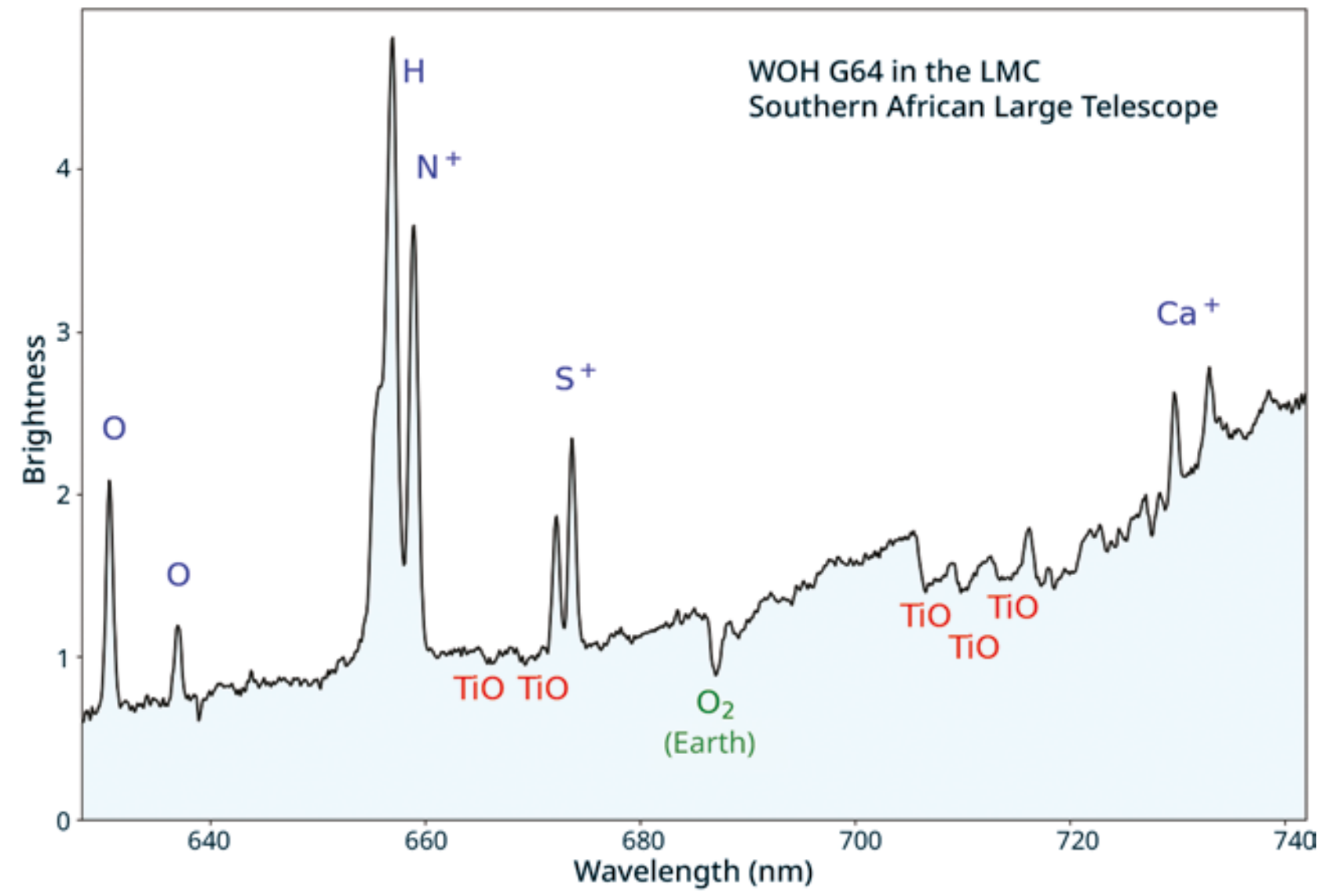

2 The red portion of the SALT spectrum of WOH G64 (van Loon & Ohnaka 2026), plotted here, shows emission spikes (in blue) from atoms caused by the hot star but also absorption dips (in red) from TiO molecules in the extended atmosphere of the red supergiant. The system is enveloped in a shroud of dust that formed in the red supergiant wind and that makes it more difficult to see the stars themselves. The absorption dip (in green) from oxygen molecules is caused by the Earth's atmosphere through which we look out into space.

variations may still be explained by pulsation even now, though we also presented clear evidence of time varying obscuration by dust. A super-expansion of the atmosphere would have enhanced the gas density at the dust condensation point. This may explain the fresh dust cloud seen in the VLTI image (Ohnaka *et al.* 2024), which was also highly flattened lending further credence to the binary interaction hypothesis, because it reflects the system's symmetry. This dust cloud was responsible for the fading of WOH G64 A, making WOH G64 B stand out more. The emission lines we attribute to component B do show signs of variable attenuation as well, which could be arise from the passage of a blobby dust wind from the red supergiant.

**"We cannot rule out that the red supergiant, alone, was responsible for its own transfiguration"**

Only time – and more observations – will tell whether or not this is correct. We cannot rule out that the red supergiant, alone, was responsible for its own transfiguration. Perhaps its mantle was already so diluted that the high rate at which its mass was reduced resulted in the retreat of the optical depth unit point and the emergence of a warmer photosphere. In that case, we expect that evolution to continue, with the yellow hypergiant throwing off its cloak and the molecular absorption dissipating. If, however, the molecular absorption persists (it was in fact strengthening in our most recent SALT spectra) and the red supergiant brightens again, then the binary interaction scenario is favoured and the red supergiant lives to see another day (or possibly century, depending on the orbital period).

It must be stressed that neither of these two scenarios predict a supernova. A yellow hypergiant may live for many thousands of years, as yellow hypergiants are most definitely known to exist. Repeat interactions between the two stars might eventually strip the envelope of the red supergiant, but that, too, does not trigger core collapse. This still leaves the question open as to the fate of WOH G64, and the – possibly diverse – nature of the progenitors of type IIb supernovae, for instance (Reguitti *et al.* 2025). None of the supernova progenitor

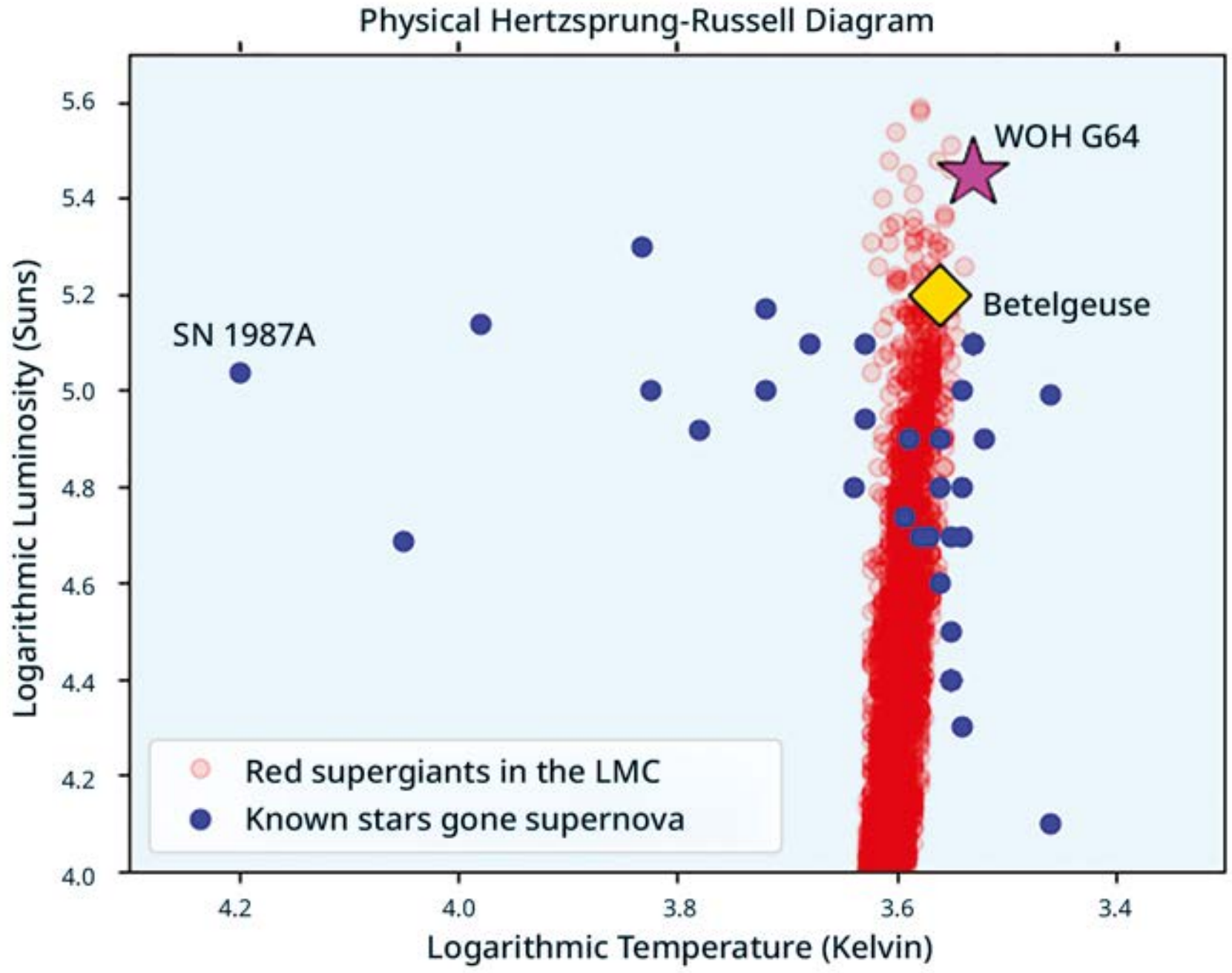

**3 When plotted in a diagram of luminosity versus temperature, red supergiants (here a sample from the LMC (Neugent *et al.* 2020)) form a slightly-inclined sequence. The stars that have been implicated in supernovae, almost all hydrogen-rich(ish), are mostly red supergiants and some yellow hypergiants at the higher luminosities (and one blue star that caused SN 1987A). Betelgeuse sits only just off the "elbow" of this "supernova alley", but WOH G64 is definitely displaced from it especially in luminosity.**

stars that have been recovered are nearly as luminous (Smartt 2015, Van Dyk 2025). WOH G64 would not have stood out in the optical images taken by the HST, but it will on infrared images taken by the JWST.

## Not every star can be an oddball

We can only call something strange if we know what normal looks like. The individual red supergiants that are studied in most detail are either the ones nearest to us, or the ones that are extreme in some way or that did something unexpected that caught our attention. The latter may be exceptional, either because of the rarity or brevity of the phase or event, or because there was something else unusual about them in the first place. Nearby examples, though, are generally expected to be typical for the population as a whole. So if Betelgeuse is normal, most red supergiants produce dust bunnies that lead to dimmings. A closer look has indeed spotted such behaviour in other red supergiants. On the other hand, SN 1987A was an unusual supernova, one kind in a hundred. Most red supergiants are therefore not expected to end that way. Where, then, does WOH G64 sit? Is it displaying ordinary behaviour, just of short duration? Or will it end like SN 1987A?

**"We can only call something strange if we know what normal looks like"**

The steepness of the initial mass function and accelerated evolution of massive stars conspire against finding red supergiants. However, when entire galaxies are surveyed their tally easily adds up to thousands. It should therefore be no surprise that a few among them turn out to be special, including WOH G64 and the progenitor of SN 1987A. Their chances of being identified increase if they were the product of a merger, since the lower-mass constituents would have been more plentiful as well as living longer. This matters because they would place the red supergiant in a less congested and confusing galactic location, aiding in discovery. As it turns out, SN 1987A is situated near the Tarantula Nebula but not in it, WOH G64 is not immersed in an active star forming complex and Betelgeuse is not near any of the Orion OB-star associations. The red supergiant progenitor of the type II-P (hydrogen-rich) SN 2018gj lived in the outskirts of the spiral galaxy NGC 6217 (Smartt 2015). Could they all have been a merger product? Mergers can be dynamically induced in triplets, which might explain the surviving companion of Betelgeuse and that of WOH G64. It could also explain the high luminosity of WOH G64 and might suggest similarly luminous red supergiants are more often the products of mergers.

To identify the work of companions we need better statistics on the incidence of binarity and higher order multiplicity among massive stars in general, and when one is a red supergiant in particular. It is also important to be able to rule out, or set limits on, the possible companionship and estimate the current mass (from the luminosity) and local star formation history (to guess the likely birth mass). This may then allow us to separate mergers from single star evolution, which is easier said than done. The high luminosity of the red supergiant easily drowns out the light from a hot companion, especially as circumstellar dust disproportionally obscures the blue light where the contrast would have been more favourable. Red supergiant photospheres reflect the dynamic behaviour of their mantle and atmosphere, making it harder to measure accurate radial velocity modulation and attributing them to orbital motion – this was also the case for Betelgeuse's long-period variability.

The outcome of binary interaction varies depending on the properties of the system (initial mass ratio, separation, eccentricity, mass loss, etc.) and could involve early mass transfer before further events unfold. Thus, the combination of their probability distributions and the exact timing of core collapse must result in a rich diversity of stellar end phases, much like the expression of interior modalities and exterior influences and interactions shape the wonderful spectrum of personalities in humans. Once the red supergiant has died, it contributes a neutron star or black hole to the graveyard and its envelope – unless fallen into the black hole – will disperse and mix with those of many stars gone before. After a tumultuous but temporary existence, stars like WOH G64 meet an everlasting, simpler fate, closer to equilibrium than during their eventful lives. In the end, entropy wins. ●

AUTHOR
**Jacco van Loon** is reader in astrophysics at Keele University and director of Keele Observatory. Their research concerns the behaviour and evolution of complex systems such as galaxies, and focuses on the process of stellar death, the chemistry and dynamics of the interstellar medium and the environment of active galactic nuclei. They have a keen interest in the physical nature of life, time and the mind, and in equity and inclusion.